\documentclass[12pt,a4paper]{article}
\usepackage[utf8]{inputenc}
\usepackage{amsmath}
\usepackage{amsfonts}
\usepackage{amssymb}
\usepackage{siunitx}

\usepackage{graphicx}
\usepackage{authblk}

\newcommand{\gstfull}{Ge$_2$Sb$_2$Te$_5$~}
\newcommand{\gst}{GST~}
\newcommand{\gstns}{GST} 
\newcommand{\sbte}{Sb$_2$Te$_3$~}

\newcommand{\sbs}{Sb$_2$S$_3$~}

\newcommand{\sio}{SiO$_2$~}

\newcommand{\znssio}{(ZnS)$_4$(SiO$_2$)$_1$~}
\newcommand{\znssions}{(ZnS)$_4$(SiO$_2$)$_1$}

\author[1]{Simon Wredh*}
\author[2]{Yunzheng Wang}
\author[1]{Joel K.W. Yang}
\author[3]{Robert E. Simpson*}
\affil[1]{Singapore University of Technology and Design (SUTD), 8 Somapah Rd, Singapore 487372}
\affil[2]{Center for Optics Research and Engineering, Shandong University, Qingdao 266237, P. R. China}
\affil[3]{School of Engineering, University of Birmingham, B15 2TT, UK}

\title{Multi-level Optical Switching by Amorphization in Single- and Multi- Phase Change Material Structures} 

\begin{document}
\maketitle


\begin{abstract}
The optical properties of phase-change materials (PCM) can be tuned to multiple levels by controlling the transition between their amorphous and crystalline phases.
In multi-material PCM structures, the number of discrete reflectance levels can be increased according to the number of PCM layers. However, the effect of increasing number of layers on quenching and reversibility has not been thoroughly studied.
In this work, the phase-change physics and thermal conditions required for reversible switching of single and multi-material PCM switches are discussed based on thermo-optical phase-change models and laser switching experiments. 
By using nanosecond laser pulses, 16 different reflectance levels in \gstfull are demonstrated via amorphization. Furthermore, a multi-material switch based on \gstfull and GeTe with four discrete reflectance levels is experimentally proven with a reversible multi-level response.
The results and design principles presented herein will impact active photonics applications that rely on dynamic multi-level operation, such as optical computing, beam steering, and next-generation display technologies.
\end{abstract}

\section{Introduction}

Materials with reconfigurable optical levels will be a key enabler of computational photonics, beam steering, and holographic displays \cite{RN5090}.
One class of material that promises multi-level operation are phase-change materials (PCM), whose programmable optical states have been technologically relevant since the initial discovery of reversible threshold switching in the 1960s \cite{ovshinskyReversibleElectricalSwitching1968}.PCMs have been particularly useful for binary data-storage applications, such as optical disks and electronic phase-change random access memory \cite{wuttigPhasechangeMaterialsRewriteable2007}.
Their success as memory materials stems from the large property contrast between crystalline and amorphous phases, fast and reversible switching, low-energy retention, non-volatile memory states, and high scalability \cite{burrPhaseChangeMemory2010, wongPhaseChangeMemory2010, raouxPhaseChangeMaterials2014}.
The large optical contrast is demonstrated in \textbf{Figure \ref{fig:single_material}a}, where a dramatic increase in reflectivity is observed as the PCM \gstfull (\gst) is heated from its amorphous state to above its crystallization temperature.
By controlling the conditions of the crystallization, or the reverse amorphisation process, multiple optical levels can be achieved, thus enabling a range of applications beyond binary data storage.

Recently, a lot of interest has been directed towards PCMs for controlling interconnections in photonic computing platforms \cite{ riosIntegratedAllphotonicNonvolatile2015,wuttigPhasechangeMaterialsNonvolatile2017,riosInmemoryComputingPhotonic2019,lianPhotonicComputationalMemories2022}.
The conventional computing paradigm of binary encoding and electronic processing limits the data storage density, processing speeds, and  energy efficiency due to unwanted Joule heating.
Using light for computing is a promising avenue to decrease latency and increase bandwidth, whilst also decreasing the energy use\cite{caulfieldWhyFutureSupercomputing2010, shenSiliconPhotonicsExtreme2019, shastriPhotonicsArtificialIntelligence2021, huangProspectsApplicationsPhotonic2022}.
In optical computing, memories that can be reconfigured into multiple optical levels simultaneously as the data is being processed, can be used to set the weights in an optical neural network \cite{RN5111}.
To satisfy the requirements of variable network weights and high density data storage of these architectures, the focus of memory design has shifted from the binary realm toward analogue and multi-level devices \cite{RN5111, wuAnalogOpticalComputing2022, kariOpticalElectricalMemories2023, huangProspectsApplicationsPhotonic2022, pitchappaChalcogenidePhaseChange2019}.
Using different degrees of crystallization of phase-change materials, is promising for programming multiple optical responses into the same material \cite{RN5042,RN1129,RN5185, RN5111}.
However, being able to reliably and precisely set synaptic weights is challenging due to noise susceptibility and variability in intermediate material states \cite{RN5087,liResistanceDriftPhase2012}.
To overcome these issues, multi-layer PCM structures with a multi-level property response could be utilised \cite{RN4807,RN4958}.
Herein, we focus our discussion around re-configurable and multi-level optical memories, which also find use in several other photonics applications, such as integrated optics, active metamaterials, and nanodisplays \cite{RN5090,wuttigPhasechangeMaterialsNonvolatile2017, wangOpticallyReconfigurableMetasurfaces2016, riosColorDepthModulation2016}.

Multi-level PCM memories can be categorised into two types: single- and multi-material systems.
In the former, the intrinsic property contrast between amorphous and crystalline states of a single material, such as \gst, has traditionally been used to store binary data. 
However, by finely controlling the level of phase-change, a gradual change in optical response can be achieved, which enables access to multiple optical property levels.
For instance, eight different reflectivity levels in \gst by picosecond laser pulses has been demonstrated \cite{riosIntegratedAllphotonicNonvolatile2015}.
Conversely, in a multi-material system, the optical response can be stored discretely in different layers. 
Different types of multi-level structures have been shown in the electrical \cite{gyanathanMultilevelPhaseChange2011, huMultiStepResistanceMemory2012, zhaoSbSeZnSbStacked2019} and optical domains \cite{mengDesign4levelActive2018}. 
We previously showed how VO$_2$ and \gst, which are two very different phase change materials, can be combined to create four discrete programmable reflectivity levels at a fixed wavelength\cite{mengDesign4levelActive2018}, and to create a mid-infrared notch filter that could block four different wavelengths\cite{RN4958}.
Vermeulen et al. demonstrated color switching in a stack consisting of three PCM layers by utilising thin-film resonance effects\cite{vermeulenMultilevelReflectanceSwitching2019}. 
In all of these previous demonstrations, however, reversibility was not demonstrated. 

We hypothesise that a multi-layer structure should be more robust than a single layer PCM structure for programming into multiple discreate optical states.
This hypothesis is based on the fact that variations in the optical response of PCMs are inevitable due to the stochastic nature of crystallisation and recrystallisation, which can lead to readout noise.
By using multiple different  PCM layers with well-separated transition temperatures, however,  the different optical states can be individually addressed, which will lead to a more reliable readout.
While the single material system maybe akin to a true analogue memory, with a continuum of programmable levels, precise control of the crystallization process is challenging \cite{RN5049}.

Although others have shown multi-level optical response from multi-layer phase change structures\cite{mengDesign4levelActive2018, vermeulenMultilevelReflectanceSwitching2019, RN4958}, reversible switching has not been demonstrated. 
The objective of this work is, therefore, to understand the factors that make reversible multi-level switching of PCM  difficult, and then to design a structure that can be reversibly switched.

We begin our discussion on multi-level optical response by studying limitations of resetting a single layer phase change material from its fully crystallised state into different levels of disorder. 
I.e. we aim to set different optical levels via the amorphization process, rather than the more typical crystallisation process. 
We study the amorphisation process by laser switching and using a Gillespie cellular automata model to predict the level of crystallinity for different quench rates.
In the second part of this manuscript, we discuss the design of a multi-material optical switch based on \gst and GeTe. 
Here, we optically and thermally engineer a PCM stack with a discrete set of optically contrasting levels that can be reversibly switched between.
These experiments and models will allow us to understand and compare the factors that influence the multi-optical-level reset process in single layer and multilayer PCM structures.

\section{Intrinsic multi-level behaviour of \gst}

Although difficult, it is possible to write a spectrum of optical levels into a single-layer \gst film.
In \gstns, different optical levels are possible due to the intrinsic complexity of bonding states that occur when amorphous \gst nucleates \cite{RN4200,RN5022} or due to partial crystal growth within an amorphised film.
\gstns, in particular, is a nucleation dominated PCM, which means that a high density of crystal nucleation sites form during crystallization. Since crystal growth is initiated from each nucleated crystallite, this high nucleation density typically means that the material rapidly reaches a fully crystallised state.
Therefore, to partially crystallise \gstns, the heat pulse must be controlled so that the crystal nuclei density is limited and controlled growth can occur from a low density of nuclei.
Since the steady state nucleation rate in \gst is greatest in the 250~C to 430~C temperature range and since it lowers logarithmically outside this range \cite{RN5186}, the time that \gst is held between 250~C and 430~C controls the density of nucleation centres, which in turn can enable a partially crystallised film with crystallites that have grown from these nucleation centres.

For PCMs on the GeTe-\sbte pseudobinary tie-line, the nucleation time is on the order of a nanosecond \cite{RN4449}, hence sub-nanosecond laser or electrical heating pulses are typically required to achieve the quench rates that can lead to partially crystallised films with concomitantly different optical properties \cite{RN5099}.
Note, some optical devices have been developed that achieve multi-level optical response by varying the length or area of the \gst film that is crystallised within an optically probed region \cite{liangEnhancedSurfaceEffects2023}, and we refer to this case as extrinsic crystallization.
Herein, however, we aim to achieve multiple optical levels by controlling the size of nanocrystalites within an amorphous matrix. 
The resultant optical response is then an intrinsic property of the resultant material. 
Control of the intrinsic optical properties through partial crystallization is particularly important for nanophotonic applications where the structural dimensions are significantly smaller than the wavelength of light. 
In this sub-wavelength regime, light no longer experiences scattering, and the optical response  depends solely on the average intrinsic properties of the material, defined by effective medium theory.

Access to different intrinsic optical levels in PCMs is typically achieved through either gradual solid-state crystallization of the amorphous phase\cite{RN1803}, or by melt-quench partial recrystallisation\cite{RN1518, RN5042}.

Typically, the electrical or optical pulses used for melt-quench partial recrystallisation of substantially greater than the minimum power needed to amorphise the PCM. 
The degree of crystallinity is then controlled by the quench rate, which is in turn controlled by the amount of additional heat deposited into the PCM and its surrounding structure.
Partial amorphisation is rarely demonstrated in the literature; i.e.  amporphising the crystalline material into different optical levels with heat pulse powers close to the threshold for complete amorphisation. 
Partial amorphisation is distinguished from melt-quenched partial recrystallisation by the behaviour of the optical reflectivity or electrical resistivity with switching energy.
Melt-queched recrystallisation occurs when the power used to switch the PCM is significantly greater than the minimum power needed to fully amorphise the material.
In this regime, for pulse amplitudes above that required for amorphisation, the crystallinity  increases with the switching power-- see the switching ``U'' curves in the works by others\cite{RN1518, RN5042}.
In contrast partial amorphisation occurs when the  power used to switch the material is slightly less than the minimum power for complete amorphisation.   
In the partial amorphisation regime the crystallinity $decreases$ with increasing switching power. 
For efficiently switching devices, it is desirable to program from either crystalline, amorphous, or intermediate states without doing a full melt-quench reset, thus enabling faster operation, lower energy consumption, and longer lifespan of devices. 
Therefore, herein, we focus on the less-studied multi-level operation by partial amorphization.

\begin{figure}
    \centering
    \includegraphics[width=\linewidth]{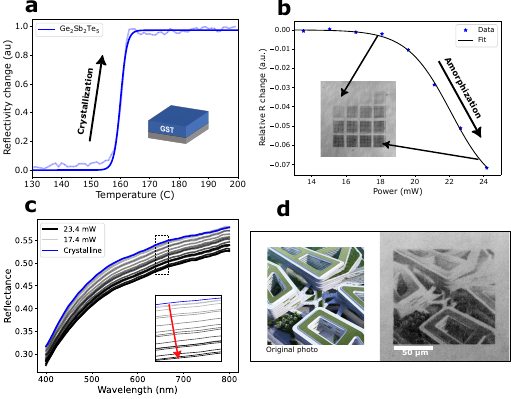}
    \caption{\textbf{(a)} Crystallization curve of \gstns. When amorphous \gst is heated above its crystallization temperature, a drastic increase in reflectance is observed. \textbf{(b)} Gradual decrease in reflectance as crystalline \gst is laser amorphized with increasing laser powers. \textbf{(c)} Recrystallization curves of amorphized GST reveals multiple optical levels. \textbf{(d)} 4-bit image written into crystalline \gst by laser-amorphization (SUTD campus from above, image credit to UNStudio).}
    \label{fig:single_material}
\end{figure}

Amorphization requires quenching the molten \gst film from a temperature above 430~$^\circ$ C, where the nucleation rates are low, and where cooling rates are sufficient to prevent the \gst film from spending a significant time at temperatures between 250~C and 430~C. 
In this case, the crystallinity after melt-quenching depends on the quench rate across the nucleation temperature range, which in the case of optical switching can be controlled by the laser fluence and the thermal properties of the PCM and surrounding materials.
Experimentally, we have used this technique to demonstrate 16 different optical levels in \gstns, by exciting a crystalline sample with nanosecond laser pulses of varying intensity.
\textbf{Figure \ref{fig:single_material}d} shows a 4-bit image that was laser amorphised into a 20 nm thick crystalline \gst film by rastering 100~ns pulses with peak optical powers controlled from 17.7 mW to 23.4 mW.
\textbf{Figure \ref{fig:single_material}b} shows the relative decrease in reflectance as the sample is laser-amorphized with increasing laser powers and \textbf{Figure \ref{fig:single_material}c} shows the reflectance spectra for 16 partially amorphised regions, which confirms the 16 different levels.
Importantly, Figure \ref{fig:single_material}b \& c also show that the \gst becomes less reflecting (i.e. more amorphous) as the laser pulse power increases. 
This behaviour is indicative of the desired partial amorphisation effect, which is very different from melt-quench partial recrystallisation, which is more commonly reported in the literature\cite{RN1518, RN5042}.

We attempt to model this partial amorphization process  using a Gillespie cellular automata (GCA) phase-change model.
Our GCA model, proposed in \cite{RN5049} and open to the interested reader on GitHub, is a cross-scale, high-precision multi-physics model, that contains laser absorption, thermal conduction, lattice micro-structure evolution, effective medium theory and Fresnel equations.

\begin{figure}
  \centering
  \includegraphics[scale=0.65, angle=90]{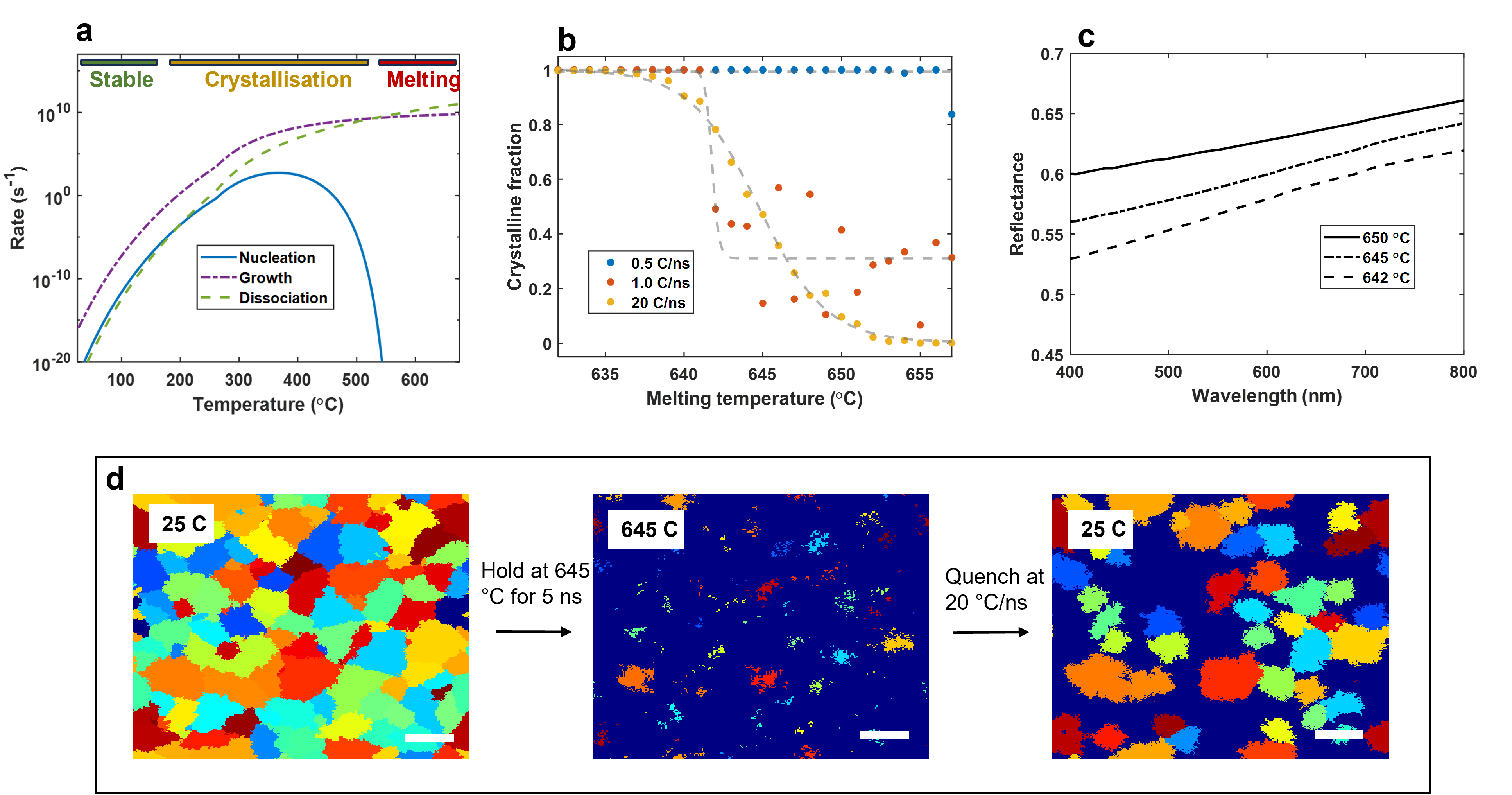}
  \caption{Partial amorphisation in \gst demonstrated by a phase-change cellular automata model. \textbf{(a)} Temperature dependent rates for nucleation, crystal growth and dissociation which governs the dynamics of the GCA model. \textbf{(b)} Partial amorphization by holding a polycrystalline film at different melting temperatures for 5 nanoseconds before quneching to room temperature. \textbf{(c)} Reflectance predictions for different holding temperatures from (b). \textbf{(d)} Left: Polycrystalline film at room temperature. Middle: Partially molten film after 5 nanoseconds at a temperature above melting. Right: Partially amorphized film after quenching the partially molten film to room temperature. Scale bars: 30 nm.}
  \label{fig:GCA_amorphization}
\end{figure}

Herein, we demonstrate that the GCA model can predict the multi-level amorphization response in \gstns. The rates for crystal nucleation, growth and dissociation of \gst are shown in \textbf{Figure \ref{fig:GCA_amorphization}a}, where dissociation can be treated as the reverse process of crystal growth that controls how crystal grains start to break down. For details on how the rates are calculated we refer to the supplementary information and reference \cite{wangSchemeSimulatingMultilevel2021}. These rates together control the crystallization dynamics of the model, and we can see that for low temperatures, the conditions are stable since all rates are low. At intermediate temperatures is where crystallization occurs due to the large rates of both nucleation and growth. At high temperatures we note that the dissociation rate overtakes the growth rate, which means that the effective growth rate turns negative and we essentially have melting. Our goal is to utilize the temperature region just above this dissociation-growth crossover-point to finely control the degree of crystal dissociation. In essence, we want to partially melt the film before quenching it back to room temperature, to achieve partial amorphization. 

First, we prepare a polycrystalline film of \gst by heating a fully amorphous simulation domain above the crystallization temperature. Next we instantaneously heat the crystalline film to a temperature just above the melting temperature and hold it there for five nanoseconds. This produces a partially molten film due to crystal dissociation, as shown in the middle panel of \textbf{Figure \ref{fig:GCA_amorphization}d}. To freeze-in the partially molten state, the film is rapidly quenched back to room temperature, and depending on the quench rate, some degree of partial recrystallization will occur during cooling, as shown in the right panel of \textbf{Figure \ref{fig:GCA_amorphization}d}.

Due to increasing dissociation rates, higher melting temperatures result in an increasing degree of crystal breakdown (melting). 
This is observed in \textbf{Figure \ref{fig:GCA_amorphization}b}, where the \SI{20}{\celsius/\ns} curve shows that good control of the remaining crystalline fraction at room temperature can be achieved by controlling the melting temperature. If we consider melting temperature as a proxy for laser power, this finding agrees well with the trend of our experimental results, which showed that increasing laser power increases the degree of amorphization.
However, to achieve fine control of partial amorphization, the rate of cooling must also be carefully considered, due to lower quench rates resulting in more recrystallization during cooling. 
For quench rates of \SI{1}{\celsius/\ns} we see a large variability in the response, and this is because the crystallites have more time to dissociate at high temperatures and more time to spontaneously nucleate and grow during cooling. 
For an even lower quench rate of \SI{0.5}{\celsius/\ns}, the film fully recrystallizes during cooling.
These results emphasize the importance of the thermal design of PCM switches.

By relating the crystalline fraction to an effective refractive index, the optical properties of the material can be predicted. This is demonstrated in \textbf{Figure \ref{fig:GCA_amorphization}c}, where the reflectance spectra has been calculated for three partially amorphized \gst films for varying melting temperatures and a quench rate of \SI{20}{\celsius/\ns}.
Based on these findings, we conclude that our GCA model, which has with crystallization physics rooted in classical nucleation and growth theory, is a powerful tool for predicting the dynamics of crystal microstructures and the optical response of PCM devices.

Thus far we have demonstrated multiple optical levels by laser amorphization in a single PCM film. 
However, programming a single layer into one of many possible close-spaced optical states comes with repeatability and reliability issues due to relaxation of the structure \cite{liResistanceDriftPhase2012, leeDistributionNanoscaleNuclei2014, RN5111}.
Indeed, others have used the randomness --- associated with the high degree of entropy in the amorphous state --- in the electrical properties of PCMs to create physically uncloneable functions \cite{goUltrafastNearIdealPhaseChange2022}.
In the following sections, we discuss how these issues of single-material devices can be overcome by designing thin film stacks with different PCMs, each having their own discrete set of optical levels.

\section{Multi-material, multi-level switch}

Many of the most useful PCMs, which have already been commercially proven in other data storage technologies, exist along the GeTe-\sbte composition tie-line\cite{RN4515}.
We hypothesized that these materials, with different crystallization temperatures, could be combined in a layered structure to form a multi-level optical switch that can be switched in nanoseconds.
The number of discrete levels would be given by $2^n$, where $n$ is the number of PCM-layers in the stack.
However, there are several challenges involved with the design of a multi-layer phase-change memory structure.
As mentioned, one difficulty with this family of PCMs is that they need to be quenched at high rates in order to form the amorphous state \cite{RN5186}. 
A multi-layer structure will make it challenging for the layers at the top of the stack, which are far from the heat sinking substrate, to quench at sufficiently high rates for amorphization.
Another difficulty is that when these materials undergo  melt-quench amorphization, they are likely to inter-diffuse. 
For this reason,  an additional layer that acts as a diffusion barrier yet minimally affects the device performance must be added between the different phase change layers.
A further consideration is the optical design. 
There must be a considerable optical contrast between each of the different optical levels for them to be distinguishable, and this depends on interlayer interference of the Fresnel equations.
Therefore, here, we needed to design a multi-layer multi-material stack that could meet these strict design criteria.

With the above considerations in mind, we sought to design a two-material structure with two main objectives. 
Firstly, the stack had to possess four discrete optical levels. 
Secondly, reversible switching should be possible between the different levels using high intensity laser pulses.

\subsection*{Design and optical contrast optimisation}
\begin{figure}
    \centering
    \includegraphics[width=\linewidth]{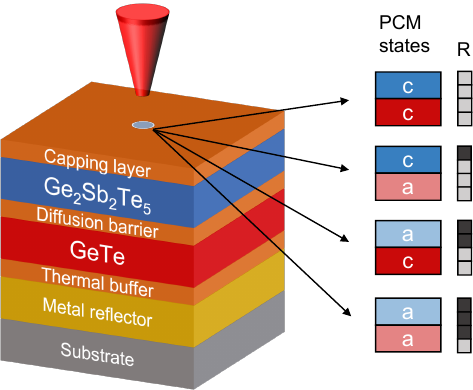}
    \caption{Two layer structure that can be programmed into four different reflectivity levels by switching each of the PCM layers between amorphous (a) and crystalline (c) states.}
    \label{fig:general_structure}
\end{figure}

The most general multi-layer structure we considered for our design is shown in \textbf{Figure \ref{fig:general_structure}}, where we chose to limit our design to two switchable layers, consisting of \gst and GeTe. 
These materials were chosen due to being well studied PCMs, yet having very different crystallization temperatures.
Supplementary Figure 1, compares the crystallization temperature of \gst and GeTe when heated at \SI{10}{\celsius/\min}. 
\gst crystallises at \SI{160}{\celsius}, whilst GeTe crystallises at \SI{234}{\celsius}.

The structure also included a capping layer, a diffusion barrier layer, and a thermal buffer layer.
The purpose of the capping layer was to protect the structure from oxidation, and the diffusion barrier was included to prevent intermixing between the GeTe and \gst layers. 
A thermal buffer layer was placed below the bottom PCM to help control the heating and quenching of the stack. 
For these three layers, we used \znssions, which is a commonly used dielectric in PCM optical data storage discs \cite{tsuMechanismPropertiesNoble2006, kuwaharaComplexRefractiveIndex2013}.
An optional metal reflector at the bottom of the stack was used to enhance the optical contrast of the different levels, and here we used a layer of Au.
Finally, to enable the required heating and high quench rates, a substrate with a suitable thermal conductivity had to be chosen, for which we compared both Si (\SI{150}{\watt/\milli\kelvin}) and \sio (\SI{1.4}{\watt/\milli\kelvin}).
The thermal conductivity of the substrates would turn out to have a significant impact on the structures' reversible switching performance.

The stack was designed with \gst on the top because it is more readily heated to its lower melting temperature (\SI{627}{\celsius}) compared to GeTe (\SI{717}{\celsius}) \cite{lankhorstModellingGlassTransition2002}.
We also considered that the quench rate of the top layer is limited by the thermal conductivity of the lower layers. 
Hence, the lower melting temperature of \gst means that there is a smaller thermal load that needs to be conducted into the substrate, and therefore higher quench rates should be possible.

\begin{figure}
    \centering 
    \includegraphics[width=\linewidth]{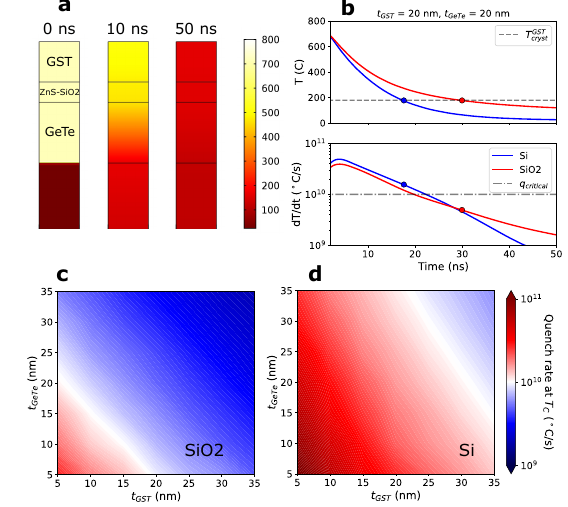}
    \caption{Quenching simulations from fully molten \gstns-\znssions-GeTe structures. \textbf{(a)} Temperature distribution snapshots. \textbf{(b)} Surface temperatures and quench rates of the top \gst layer in a 20~nm-10~nm-20~nm structure for different substrates. The colored markers indicate the points in the curves where the crystallization temperature has been reached. \textbf{(c, d)} Instantaneous quench rates for different combinations of PCM layer thicknesses.}
    \label{fig:thermal_sim}
\end{figure}

The thicknesses of the layers has to be constrained to allow the top layer to quench at sufficient rates through the rest of the stack. Amorphization of a fully molten top layer can be ensured by maintaining a quench rate above a critical value until the crystallization temperature is reached, and by analyzing the cooling dynamics of different multi-layer structures using a heat equation solver, appropriate thickness constraints can be determined.
We did this by first instantaneously heating the layers of a \gstns-\znssio-GeTe stack to the melting temperature of GeTe (\SI{717}{\celsius}), whilst keeping the initial temperature of the substrate at room temperature. 
We then modelled how the heat dissipated through the structure as it cooled, and from the temperature distribution snapshots in \textbf{Figure \ref{fig:thermal_sim}a} we can see that close-to-room temperature is reached after 50 nanoseconds.
We further monitored the surface temperature and the instantaneous quench rate of the structure for both Si and \sio substrates, and \textbf{Figure \ref{fig:thermal_sim}b} shows these quantities as a function of time for a \gst and GeTe thickness of 20~nm each, separated by a 10~nm \znssio layer.
We notice, as expected, that the Si substrate with its higher thermal conductivity allows the top layer to quench to the crystallization temperature faster than \sio. From the GCA simulations in \textbf{Figure \ref{fig:GCA_amorphization}}, we can reason that a quench rate above a critical value of \SI{e10}{\celsius/\second} should be maintained until the crystallization temperature is reached to fully amorphize the top layer, and by studying the instantaneous quench rate curves, we realize that this is achieved for Si, but not \sio. 
This emphasizes the importance of the thermal design to ensure reversible switching. 
To understand which stack configurations would allow full amorphization, we ran the model for different combination of PCM layer thicknesses, and plotted the instantaneous quench rate at the crystallization temperature in \textbf{Figure \ref{fig:thermal_sim}c} and \textbf{Figure \ref{fig:thermal_sim}d} for Si and \sio substrates. From these plots it becomes apparent that a higher thermally conductive substrate allows for more flexibility in PCM-layer thickness when it comes to ensuring sufficient quench rates for amorphization.
Based on these modelling results, which assumes that the whole structure can be molten, we chose to constrain the \gst and GeTe thicknesses to 30~nm and 25~nm for Si, and 15~nm and 15~nm for \sio substrates.

\begin{figure}
    \centering
    \includegraphics[width=\linewidth]{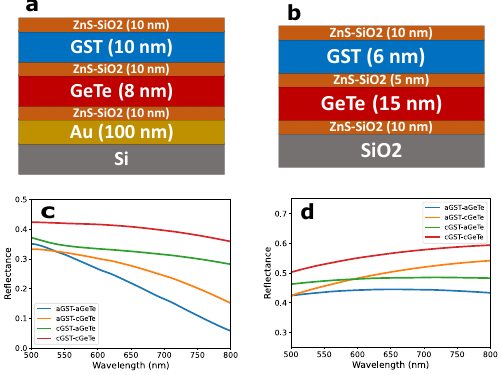}
    \caption{Diagrams of the fabricated multi-level structures, with corresponding theoretical reflectance spectra of the four levels, on Si (\textbf{a, c}) and SiO2 \textbf{(b, d)} substrates}
    \label{fig:specific_structure}
\end{figure}

Besides influencing the thermal conditions required for phase-change, choosing the correct layer thicknesses is also crucially important to ensure a good optical design. Indeed, the thicknesses of each layer in the stack must be optimised to maximise the contrast between the different PCM-state combinations, which are aGST-aGeTe, aGST-cGeTe, cGST-aGeTe and cGST-cGeTe.
We searched for an optimal structure by combining a transfer matrix solver of the multi-layered Fresnel equations with a genetic search algorithm. 
Given a structure with a specified sequence of layers and corresponding thickness constraints, the search routine would find the combination of layer thicknesses that resulted in the largest contrast between each neighbouring reflectance level.
The code for this optimizer is available on GitHub \cite{Wredh_Contrast_optimisation_for_2023}.
By using this approach, we designed two structures that we could prototype and characterize.
These structures and their theoretical reflectance spectra are shown in \textbf{Figure \ref{fig:specific_structure}}.
The structures were optimized to achieve a maximum contrast at a wavelength of 650~nm to match the wavelength of the probe laser in our laser switching system.

\subsection*{Experiments}

To evaluate the two prototype structures, we fabricated them by sputter deposition, and tested the reflectance levels during furnace annealing and nanosecond laser switching.

\begin{figure}
    \centering
    \includegraphics[width=\linewidth]{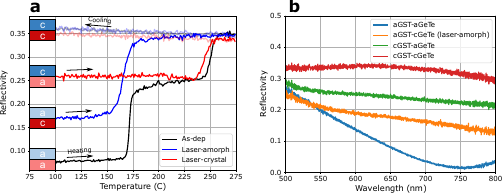}
    \caption{\textbf{(a)} Crystallization curves during furnace heating for different initial states of the Si stack. The transparent curves show the cooling process. \textbf{(b)} Measured reflectance spectra of the four levels in (a).}
    \label{fig:silicon_substrate}
\end{figure}

Due to the thinness of the layers and the fact that the optical response is very sensitive to the layer thicknesses, it is important to verify the integrity of the layers and confirm that the designed stack was fabricated successfully. 
Since the expected optical response is known for each PCM state, the as-deposited structure, which is fully amorphous, can be validated by heating it across the crystallization temperatures of both PCMs to fully crystallize the stack while measuring its reflectance.
We did this for the Si stack and its reflectance (at $\lambda=650$~nm) as a function of temperature is shown in \textbf{Figure \ref{fig:silicon_substrate}a}. 
The black curve shows the heating curve for the as-deposited structure, and the two distinct events at \SI{170}{\celsius} and \SI{250}{\celsius} confirm that contrasting reflectance levels exist in the stack and that no intermixing between \gst and GeTe has occured.
Furthermore, this demonstrates that three of the expected reflectance levels exist in the stack: aGST-aGeTe, cGST-aGeTe and cGST-cGeTe.
To demonstrate the fourth level, aGST-cGeTe, we laser-amorphized the top \gst layer of a fully crystalline sample, and then repeated the heating experiment.
This resulted in the blue curve in \textbf{Figure \ref{fig:silicon_substrate}a}, which only displays the \gst crystallization event, thus confirming the fourth level.
Similarly, the red curve in the same plot shows a single GeTe crystallization event, from an amorphous sample where only the top \gst layer had been laser crystallized.
The top left corner of the plot shows the reflectance level of the cooled and fully crystalline stack, and thus all four discrete levels have been demonstrated at room temperature.
\textbf{Figure \ref{fig:silicon_substrate}b} shows the reflectance spectra for each level over a broader wavelength range, which is in fairly good agreement with the theoretical spectra from \textbf{Figure \ref{fig:specific_structure}c}. From these crystallization experiments we can thus conclude that the designed structure was successfully fabricated with four discrete reflectance levels. 

\begin{figure}
    \centering
    \includegraphics[width=\linewidth]{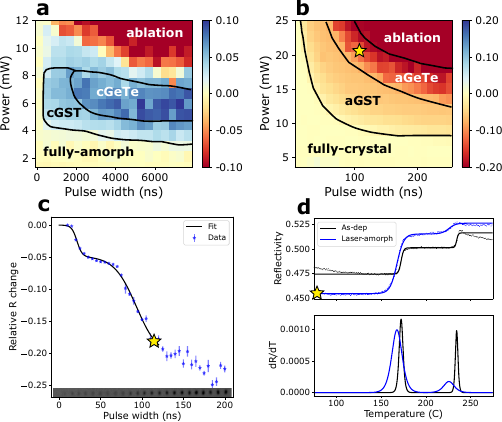}
    \caption{Power-time reflectivity plots for crystallization (a) and amorphization (b) of the SiO2 stack. \textbf{(c)} Reflectivity change due to amorphization pulses of increasing duration, sliced from (b) at 21 mW. The second transition around 100 ns suggests successful amorphization of the GeTe layer. \textbf{(d)} The two transition events in the recrystallization curve confirms that both the GST and GeTe layers were amorphized.}
    \label{fig:silica_substrate}
\end{figure}

To study the switchability of the structures, we used high intensity laser pulses to switch the PCM layers from both amorphous and crystalline state, while measuring the relative change in reflectance. By varying the power and the pulse width of the laser pulses, a power-time-reflectance (PTR) contour can be obtained which characterizes the switching properties of a given initial state.
We show the PTR plots for the \sio structure In \textbf{Figure \ref{fig:silica_substrate}a} for crystallization of the fully amorphous stack, and in \ref{fig:silica_substrate}b for amorphization of the fully crystalline stack. In these plots, a positive change of reflectance is shown in blue and indicates crystallization, whereas a negative change shown in red indicates amorphization, or for higher energies ablation.
We first note that longer pulses of lower power are used for crystallization, whereas shorter and higher powered pulses are required for amorphization, which is what we typically expect for PCMs.
We can further observe discrete steps in the reflectance change for increasing pulse energies, as indicated by the different shades of blue and red.
This result suggests that we are able to access each of the four levels and that the stack can be reversibly switched. 
The laser accessed levels are highlighted in \textbf{Figure \ref{fig:silica_substrate}c}, which shows a constant power slice from the amorphization PTR plot at 21~mW, and here the separate onset of amorphization of the two materials can be observed: for \gst between \SI{20}{\ns} and \SI{30}{\ns}, and for both \gst and GeTe between \SI{75}{\ns} and \SI{125}{\ns}.
It should be noted, however, that the simultaneous amorphization of the \gst and GeTe happens for pulse parameters very close to the energies that cause ablation damage in the top \gst layer. This suggest that the thermal conductivity of the \sio substrate may be too low, causing the upper layers to get too hot during laser irradiation. Conversely, for the stack with the Si substrate, the highest laser energies applied were unable to induce full amorphization, thus suggesting that an optimal substrate should have a thermal conductivity in-between the two substrates considered here.

To confirm that the structure could be fully amorphized and thus reversibly switched, we irradiated a large area using the laser parameters indicated by the star in \textbf{Figure \ref{fig:silica_substrate}b}, which we then recrystallized using the heating stage. 
The resulting recrystallization curve (blue) is shown together with the as-deposited curve (black) in \textbf{Figure \ref{fig:silica_substrate}d}. 
Additionally, a double logistic function was fit to the data and differentiated to identify any crystallization events. 
In the bottom, differentiated plot, we can observe two peaks in the recrystallization curve that match with the as-deposited crystallization peaks for \gst and GeTe. 
Although, the GeTe peak at \SI{225}{\celsius} is smaller than that of the as-deposited structure, it confirms that the bottom layer was at least partially amorphised, which demonstrates reversibility.
We note that the reflectance of the laser-amorphised structure is lower than the as-deposited state. This effect is likely due to diffractive losses introduced by a minor surface roughening of the top surface, caused by the high powered laser irradiation.
The slight shift to lower crystallization temperatures for the amorphised structure is due to the presence of crystal nuclei that remain after partial laser-amorphization, as illustrated in \textbf{Figure \ref{fig:GCA_amorphization}d}. During recrystallization, these nuclei act as seeds for earlier crystal growth to occur \cite{weiTheoreticalExplanationDifferent2003, raoux2008crystallization}.

\begin{figure}
    \centering
    \includegraphics[width=\linewidth]{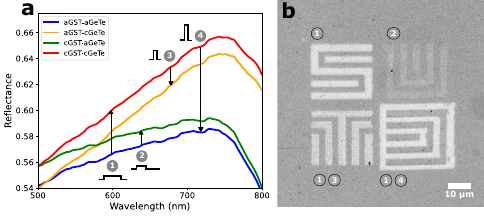}
    \caption{\textbf{(a)} Schematic of how each of the four reflectance levels can be accessed by laser pulses. The reflectance spectra are measured. \textbf{(b)} Demonstration of the four levels, each laser-written into one of the letters S U T D in an as-deposited sample of the \sio structure. The image which shows the red channel of the original micrograph has had its contrast enhanced to highlight the different levels.}
    \label{fig:level_access}
\end{figure}

Based on these amorphization and recrystallization studies, we firstly conclude that the multi-layer PCM structure can accommodate four discrete reflectance levels, assuming an appropriate optical design.
Secondly, reversible switching between the four levels is possible, assuming an appropriate thermal design. 
\textbf{Figure \ref{fig:level_access}a} shows a schematic on top of measured reflectance spectra for the \sio stack of how each level can be accessed, and we demonstrate the controlled switching in \textbf{Figure \ref{fig:level_access}b}, where each of the four levels have been laser written into an as-deposited (amorphous) sample.
These results highlight the possibility to optically program multiple levels in multi-layered PCM structures, which is very relevant to achieve dynamic and multi-functional optical devices for integrated photonics applications. We also envisage that these results will be used to extend the functionality of multi-layered metasurface systems, where the phase delay of the reflected or transmitted light needs to be controlled in steps across $2\pi$ radians.

Designing a multi-material multi-level photonic switch, like the one presented herein, entails a complex multi-physical optimisation problem, which requires simultaneous consideration of optical, thermal and phase-transition physics.
Attempting to design a device with the only goal to achieve optimal optical contrast may result in a structure that cannot be reversibly switched, as turned out to be the case for the Si structure in \textbf{Figure \ref{fig:specific_structure}a}.
Thus, to ensure reversible switching, great care must be given to the thermal design, where the main property of interest is the thermal conductivity of the materials, especially of the substrate, which acts as the heat-sink of the stack.
In this study we realised that the thermal conductivity of \sio (1.4~W/mK), which is two orders of magnitude lower than that of Si (150~W/mK), enabled switching of both layers.
The thermal conductivity of the substrate must be sufficiently low  for each PCM layer to be heated above their melting temperatures, but it must also be high enough to enable the necessary quench rates required for amorphization. 
This design contradiction is a well known constraint for reversible switching of single PCM films, however, the multi-material case is more complex since the constraints on different layers are less predictable.
For instance, the main constraint on the upper layer is that it must be able to dissipate heat through the lower layers fast enough to freeze-in the molten state's disordered atomic configurations.
In contrast, the bottom layer is mainly constrained by reaching a sufficiently high temperature to melt, without overheating the top layer to the point of ablation. 
However, since the bottom layer is closer to the heat sinking substrate, achieving high quench rates is less of a concern. 
For reversible switching, the number of reflectance levels cannot simply scale according to $2^n$ because of these thermal constraints.

By employing the design principles presented herein, other multi-layered configurations could be considered to achieve different switching functionalities. 
For example, transmittance and phase-shift levels could be implemented using similar designs. 
A further option is to create a wavelength sensitive switch by combining PCMs with different bandgaps, such as \gst (cryst. \SI{0.5}{\eV}, amorph. \SI{0.7}{\eV}) and \sbs (cryst. \SI{1.7}{\eV}, amorph. \SI{2.0}{\eV}) \cite{dongWideBandgapPhase2018}. 
Here, we would put \sbs on top and \gst on the bottom. 
In this design, incoming light with energies above the bandgap of \sbs would be mainly absorbed in the top layer to initiate switching, whereas light with intermediate energies between the two bandgaps would pass through the top layer and switch only the bottom \gst layer.

\section{Conclusions}
We have discussed the thermal conditions and constraints involved with reversible switching of single and multi-material PCM structures; particularly focusing on the amorphization process. 
We designed and fabricated a single and multi-material PCM switches with a discrete set of reflectance levels that can be reversibly switched  using nanosecond laser pulses. 
In the multi-material case, the thermal conductivity of the substrate was of utmost importance; too high conductivity and reversible switching is prevented, and too low can result in structural damage to the upper layers of the multi-material stack during laser irradiation.
We believe that the general design rules presented herein will impact the design of multiple optical state reconfigurable displays, metamaterials, and photonics integrated circuits.

\section{Experimental Section}

\*subsection{Sample fabrication}
All films were sputter deposited in their amorphous state by an AJA sputtering system at room temperature in an argon atmosphere. Radio-frequency guns were used to deposit GeTe (10 W) and \znssio (80 W), whilst a DC gun was used to deposit \gstns (15 W). Deposition powers in parentheses.
The sputtering chamber's base pressure was \SI{5e-7}{mTorr} before deposition and \SI{3.7}{mTorr} during deposition of each layer.
All sputtering targets had a two inch diameter.

\subsection*{Heating stage}
The temperature dependent reflectance measurements were performed in a Linkam heating stage coupled to a microscope spectrometer with a spectral range of 400-800 nm. The samples were heated from room temperature to \SI{275}{\celsius} in an argon atmosphere at a heating rate of \SI{10}{\celsius\per\min}.

\subsection*{Laser switching}
The laser-induced crystallization and amorphization was characterized using an in-house built static tester \cite{beheraLaserSwitchingCharacterisation2017}, where a 660 nm pump laser with a beam radii of 0.8 um and a 638 nm probe laser were focused at the same spot on the sample.
High energy pump pulses of varying intensity and duration were used to induce phase-change in the materials, while the low energy probe beam was used to measure the corresponding change in reflectance by reflecting off the sample into a powermeter.
The static tester was also used in combination with a linear stage to laser amorphize and crystallize large regions for measurements using the Linkam microscope.

\subsection*{Simulations}
The amorphization results for multi-level operation in a single material (Figure \ref{fig:GCA_amorphization}) were simulated using a Gillespie Cellular Automata (GCA) code \cite{RN5049}. 
The simulation domain was initialized in a fully amorphous state at the melting temperature (627~C) of \gstns, from which the system was quenched linearly at different rates to room temperature, where the remaining fraction of crystalline regions was recorded.

The thermal quenching simulations (Figure \ref{fig:thermal_sim}) were performed by solving the transient heat equation. A 2D geometry was used with periodic boundary conditions on the sides of the structure. A convective heat flux (heat transfer coefficient \SI{20}{\watt\per\meter\squared\per\kelvin}) was set at the top of the film to mimic natural air convection. A constant temperature (\SI{20}{\celsius}) boundary condition was set at the bottom of the substrate, which was sufficiently thick to ensure physical heat transfer.

\medskip
\textbf{Supporting Information} \par 
Supporting Information is available from the Wiley Online Library or from the author.

\medskip
\textbf{Acknowledgements} \par 
This research was supported by the NSLM project (A18A7b0058). 
Mr Wredh is grateful for his Singapore Ministry of Education (MoE) PhD scholarship.

\bibliographystyle{MSP}
\bibliography{multi_level.bib,robbib.bib}

\begin{thebibliography}{10}
\providecommand{\url}[1]{\texttt{#1}}
\providecommand{\urlprefix}{URL }

\bibitem{RN5090}
R.~E. Simpson, J.~K.~W. Yang, J.~Hu,
\newblock \emph{Optical Materials Express} \textbf{2022}, \emph{12}, 6 2368.

\bibitem{ovshinskyReversibleElectricalSwitching1968}
S.~R. Ovshinsky,
\newblock \emph{Physical Review Letters} \textbf{1968}, \emph{21}, 20 1450.

\bibitem{wuttigPhasechangeMaterialsRewriteable2007}
M.~Wuttig, N.~Yamada,
\newblock \emph{Nature Materials} \textbf{2007}, \emph{6}, 11 824.

\bibitem{burrPhaseChangeMemory2010}
G.~W. Burr, M.~J. Breitwisch, M.~Franceschini, D.~Garetto, K.~Gopalakrishnan,
  B.~Jackson, B.~Kurdi, C.~Lam, L.~A. Lastras, A.~Padilla, B.~Rajendran,
  S.~Raoux, R.~S. Shenoy,
\newblock \emph{Journal of Vacuum Science \& Technology B} \textbf{2010},
  \emph{28}, 2 223.

\bibitem{wongPhaseChangeMemory2010}
H.-S.~P. Wong, S.~Raoux, S.~Kim, J.~Liang, J.~P. Reifenberg, B.~Rajendran,
  M.~Asheghi, K.~E. Goodson,
\newblock \emph{Proceedings of the IEEE} \textbf{2010}, \emph{98}, 12 2201.

\bibitem{raouxPhaseChangeMaterials2014}
S.~Raoux, F.~Xiong, M.~Wuttig, E.~Pop,
\newblock \emph{MRS Bulletin} \textbf{2014}, \emph{39}, 8 703.

\bibitem{riosIntegratedAllphotonicNonvolatile2015}
C.~R{\'i}os, M.~Stegmaier, P.~Hosseini, D.~Wang, T.~Scherer, C.~D. Wright,
  H.~Bhaskaran, W.~H.~P. Pernice,
\newblock \emph{Nature Photonics} \textbf{2015}, \emph{9}, 11 725.

\bibitem{wuttigPhasechangeMaterialsNonvolatile2017}
M.~Wuttig, H.~Bhaskaran, T.~Taubner,
\newblock \emph{Nature Photonics} \textbf{2017}, \emph{11}, 8 465.

\bibitem{riosInmemoryComputingPhotonic2019}
C.~R{\'i}os, N.~Youngblood, Z.~Cheng, M.~Le~Gallo, W.~H.~P. Pernice, C.~D.
  Wright, A.~Sebastian, H.~Bhaskaran,
\newblock \emph{Science Advances} \textbf{2019}, \emph{5}, 2 eaau5759.

\bibitem{lianPhotonicComputationalMemories2022}
C.~Lian, C.~Vagionas, T.~Alexoudi, N.~Pleros, N.~Youngblood, C.~R{\'i}os,
\newblock \emph{Nanophotonics} \textbf{2022}, \emph{11}, 17 3823.

\bibitem{caulfieldWhyFutureSupercomputing2010}
H.~J. Caulfield, S.~Dolev,
\newblock \emph{Nature Photonics} \textbf{2010}, \emph{4}, 5 261.

\bibitem{shenSiliconPhotonicsExtreme2019}
Y.~Shen, X.~Meng, Q.~Cheng, S.~Rumley, N.~Abrams, A.~Gazman, E.~Manzhosov,
  M.~S. Glick, K.~Bergman,
\newblock \emph{Journal of Lightwave Technology} \textbf{2019}, \emph{37}, 2
  245.

\bibitem{shastriPhotonicsArtificialIntelligence2021}
B.~J. Shastri, A.~N. Tait, T.~{Ferreira de Lima}, W.~H.~P. Pernice,
  H.~Bhaskaran, C.~D. Wright, P.~R. Prucnal,
\newblock \emph{Nature Photonics} \textbf{2021}, \emph{15}, 2 102.

\bibitem{huangProspectsApplicationsPhotonic2022}
C.~Huang, V.~J. Sorger, M.~Miscuglio, M.~{Al-Qadasi}, A.~Mukherjee, L.~Lampe,
  M.~Nichols, A.~N. Tait, T.~{Ferreira de Lima}, B.~A. Marquez, J.~Wang,
  L.~Chrostowski, M.~P. Fok, D.~Brunner, S.~Fan, S.~Shekhar, P.~R. Prucnal,
  B.~J. Shastri,
\newblock \emph{Advances in Physics: X} \textbf{2022}, \emph{7}, 1 1981155.

\bibitem{RN5111}
T.~Y. Teo, X.~Ma, E.~Pastor, H.~Wang, J.~K. George, J.~K.~W. Yang, S.~Wall,
  M.~Miscuglio, R.~E. Simpson, V.~J. Sorger,
\newblock \emph{Nanophotonics} \textbf{2022}.

\bibitem{wuAnalogOpticalComputing2022}
J.~Wu, X.~Lin, Y.~Guo, J.~Liu, L.~Fang, S.~Jiao, Q.~Dai,
\newblock \emph{Engineering} \textbf{2022}, \emph{10} 133.

\bibitem{kariOpticalElectricalMemories2023}
S.~R. Kari, C.~A. R{\'i}os~Ocampo, L.~Jiang, J.~Meng, N.~Peserico, V.~J.
  Sorger, J.~Hu, N.~Youngblood,
\newblock \emph{IEEE Journal of Selected Topics in Quantum Electronics}
  \textbf{2023}, \emph{29}, 2: Optical Computing 1.

\bibitem{pitchappaChalcogenidePhaseChange2019}
P.~Pitchappa, A.~Kumar, S.~Prakash, H.~Jani, T.~Venkatesan, R.~Singh,
\newblock \emph{Advanced Materials} \textbf{2019}, \emph{31}, 12 1808157.

\bibitem{RN5042}
S.~R. Ovshinsky,
\newblock \emph{Japanese Journal of Applied Physics Part 1-Regular Papers Short
  Notes \& Review Papers} \textbf{2004}, \emph{43}, 7b 4695.

\bibitem{RN1129}
M.~Miscuglio, G.~C. Adam, D.~Kuzum, V.~J. Sorger,
\newblock \emph{Apl Materials} \textbf{2019}, \emph{7}, 10 100903.

\bibitem{RN5185}
S.~G. Sarwat, F.~Bruckerhoff-Pluckelmann, S.~G.~C. Carrillo, E.~Gemo,
  J.~Feldmann, H.~Bhaskaran, C.~D. Wright, W.~H.~P. Pernice, A.~Sebastian,
\newblock \emph{Science Advances} \textbf{2022}, \emph{8}, 22.

\bibitem{RN5087}
T.~Y. Teo, M.~Krbal, J.~Mistrik, J.~Prikryl, L.~Lu, R.~E. Simpson,
\newblock \emph{Optical Materials Express} \textbf{2022}, \emph{12}, 2 606.

\bibitem{liResistanceDriftPhase2012}
J.~Li, B.~Luan, C.~Lam,
\newblock In \emph{2012 {{IEEE International Reliability Physics Symposium}}
  ({{IRPS}})},
\newblock ISSN 1938-1891, \textbf{2012} 6C.1.1--6C.1.6.

\bibitem{RN4807}
Y.~Meng, J.~K. Behera, Y.~J. Ke, L.~Chew, Y.~Wang, Y.~Long, R.~E. Simpson,
\newblock \emph{Applied Physics Letters} \textbf{2018}, \emph{113}, 7 071901.

\bibitem{RN4958}
Y.~Meng, D.~Li, C.~Zhang, Y.~Wang, R.~E. Simpson, Y.~Long,
\newblock \emph{Applied Physics Letters} \textbf{2021}, \emph{119}, 14 141109.

\bibitem{wangOpticallyReconfigurableMetasurfaces2016}
Q.~Wang, E.~T.~F. Rogers, B.~Gholipour, C.-M. Wang, G.~Yuan, J.~Teng, N.~I.
  Zheludev,
\newblock \emph{Nature Photonics} \textbf{2016}, \emph{10}, 1 60.

\bibitem{riosColorDepthModulation2016}
C.~R{\'i}os, P.~Hosseini, R.~A. Taylor, H.~Bhaskaran,
\newblock \emph{Advanced Materials} \textbf{2016}, \emph{28}, 23 4720.

\bibitem{gyanathanMultilevelPhaseChange2011}
A.~Gyanathan, Y.-C. Yeo,
\newblock \emph{Journal of Applied Physics} \textbf{2011}, \emph{110}, 12
  124517.

\bibitem{huMultiStepResistanceMemory2012}
Y.~Hu, M.~Sun, S.~Song, Z.~Song, J.~Zhai,
\newblock \emph{Integrated Ferroelectrics} \textbf{2012}, \emph{140}, 1 8.

\bibitem{zhaoSbSeZnSbStacked2019}
Z.~Zhao, S.~Hua, B.~Shen, J.~Zhai, T.~Lai, S.~Song, Z.~Song,
\newblock \emph{Journal of Materials Science: Materials in Electronics}
  \textbf{2019}, \emph{30}, 16 15024.

\bibitem{mengDesign4levelActive2018}
Y.~Meng, J.~K. Behera, Y.~Ke, L.~Chew, Y.~Wang, Y.~Long, R.~E. Simpson,
\newblock \emph{Applied Physics Letters} \textbf{2018}, \emph{113}, 7 071901.

\bibitem{vermeulenMultilevelReflectanceSwitching2019}
P.~A. Vermeulen, D.~T. Yimam, M.~A. Loi, B.~J. Kooi,
\newblock \emph{Journal of Applied Physics} \textbf{2019}, \emph{125}, 19
  193105.

\bibitem{RN5049}
Y.~Wang, J.~Ning, L.~Lu, M.~Bosman, R.~E. Simpson,
\newblock \emph{npj Computational Materials} \textbf{2021}, \emph{7}, 1 183.

\bibitem{RN4200}
M.~Krbal, A.~V. Kolobov, P.~Fons, J.~Tominaga, S.~R. Elliott, J.~Hegedus,
  T.~Uruga,
\newblock \emph{Physical Review B} \textbf{2011}, \emph{83}, 5 054203.

\bibitem{RN5022}
T.~H. Lee, S.~R. Elliott,
\newblock \emph{Adv Mater} \textbf{2020}, \emph{32}, 28 e2000340.

\bibitem{RN5186}
J.~Orava, A.~L. Greer,
\newblock \emph{Acta Materialia} \textbf{2017}, \emph{139} 226.

\bibitem{RN4449}
J.~H. Coombs, A.~P. J.~M. Jongenelis, W.~Vanesspiekman, B.~A.~J. Jacobs,
\newblock \emph{Journal of Applied Physics} \textbf{1995}, \emph{78}, 8 4918.

\bibitem{RN5099}
S.~Wen, Y.~Meng, M.~Jiang, Y.~Wang,
\newblock \emph{Scientific Reports} \textbf{2018}, \emph{8}, 1 4979.

\bibitem{liangEnhancedSurfaceEffects2023}
J.~Liang, G.~Chen, X.~Niu, Z.~Zhu, Y.~Dong, Y.~Wang, Q.-H. Wei, J.~Pan, Y.~Li,
  C.~Gu, M.~Shen, X.-D. Xiang,
\newblock \emph{Optical Materials Express} \textbf{2023}, \emph{13}, 3 566.

\bibitem{RN1803}
C.~D. Wright, Y.~Liu, K.~I. Kohary, M.~M. Aziz, R.~J. Hicken,
\newblock \emph{Advanced Materials} \textbf{2011}, \emph{23}, 30 3408–3413.

\bibitem{RN1518}
J.~Siegel, A.~Schropp, J.~Solis, C.~N. Alfonso, M.~Wuttig,
\newblock \emph{Appl. Phys. Lett.} \textbf{2004}, \emph{84}, 13 2250.

\bibitem{wangSchemeSimulatingMultilevel2021}
Y.~Wang, J.~Ning, L.~Lu, M.~Bosman, R.~E. Simpson,
\newblock \emph{npj Computational Materials} \textbf{2021}, \emph{7}, 1 183.

\bibitem{leeDistributionNanoscaleNuclei2014}
B.-S. Lee, K.~Darmawikarta, S.~Raoux, Y.-H. Shih, Y.~Zhu, S.~G. Bishop, J.~R.
  Abelson,
\newblock \emph{Applied Physics Letters} \textbf{2014}, \emph{104}, 7 071907.

\bibitem{goUltrafastNearIdealPhaseChange2022}
S.-X. Go, Q.~Wang, K.~G. Lim, T.~H. Lee, N.~Bajalovic, D.~K. Loke,
\newblock \emph{Advanced Science} \textbf{2022}, \emph{9}, 36 2204453.

\bibitem{RN4515}
N.~Yamada, E.~Ohno, K.~Nishiuchi, N.~Akahira, M.~Takao,
\newblock \emph{Journal of Applied Physics} \textbf{1991}, \emph{69}, 5 2849.

\bibitem{tsuMechanismPropertiesNoble2006}
D.~V. Tsu, T.~Ohta,
\newblock \emph{Japanese Journal of Applied Physics} \textbf{2006}, \emph{45},
  8R 6294.

\bibitem{kuwaharaComplexRefractiveIndex2013}
M.~Kuwahara, O.~Suzuki, T.~Yagi, N.~Taketoshi,
\newblock \emph{Japanese Journal of Applied Physics} \textbf{2013}, \emph{52},
  12R 128003.

\bibitem{lankhorstModellingGlassTransition2002}
M.~H.~R. Lankhorst,
\newblock \emph{Journal of Non-Crystalline Solids} \textbf{2002}, \emph{297}, 2
  210.

\bibitem{Wredh_Contrast_optimisation_for_2023}
S.~Wredh,
\newblock {Contrast optimisation for multi-level multi-layer optical switch},
  \textbf{2023},
\newblock \urlprefix\url{https://github.com/simonwredh/contrast\_opt}.

\bibitem{weiTheoreticalExplanationDifferent2003}
J.~Wei, F.~Gan,
\newblock \emph{Thin Solid Films} \textbf{2003}, \emph{441}, 1 292.

\bibitem{raoux2008crystallization}
S.~Raoux, R.~Shelby, B.~Munoz, M.~Hitzbleck, D.~Krebs, M.~Salinga, M.~Woda,
  M.~Austgen, K.-M. Chung, M.~Wuttig,
\newblock In \emph{Europ. Phase Change and Ovonic Science Symp., Prague, Czech
  Republic}. \textbf{2008} .

\bibitem{dongWideBandgapPhase2018}
W.~Dong, H.~Liu, J.~K. Behera, L.~Lu, R.~J.~H. Ng, K.~V. Sreekanth, X.~Zhou,
  J.~K.~W. Yang, R.~E. Simpson,
\newblock \emph{Advanced Functional Materials} \textbf{2018}, \emph{29}, 6.

\bibitem{beheraLaserSwitchingCharacterisation2017}
J.~K. Behera, X.~Zhou, J.~Tominaga, R.~E. Simpson,
\newblock \emph{Optical Materials Express} \textbf{2017}, \emph{7}, 10.

\end{thebibliography}

\end{document}